# Remembering the Tevatron: The Machine(s)

S. D. Holmes
*Fermilab, P.O. Box 500, Batavia, IL 60510, USA*

For 25 years the Tevatron proton-antiproton collider was the highest energy collider in the world. This presentation will trace the origins of the Tevatron, the challenges that were overcome in creating high luminosity collisions of protons and antiprotons, the technological achievements that drove performance a factor of 400 beyond the initial performance goals, and the legacy of the Tevatron in paving the way for ever more advanced colliders.

## 1.  Introduction

This talk represents the personal perspective of the author. I arrived at Fermilab in June 1983, one month before the first acceleration of beam in the Tevatron. I was not directly involved in that achievement, but joined the team designing and constructing the antiproton source. Over the subsequent 28 years I worked on every accelerator in the Fermilab complex.

There is no attempt to be historically comprehensive – for a more comprehensive description of the laboratory and the accelerators see references [1, 2, 3, 4]. Rather I will present a recollection of the events and people who I view as critical to establishing the initial vision, constructing the machine, and finally pushing performance a factor of 400 beyond the initial performance goal.

## 2.  The Vision, Construction, and Initial Operations (1972-1985)

Robert R. Wilson, the founding Director of Fermilab, provided the vision. The original 500 GeV accelerator (known as the Main Ring) was based on conventional, room temperature, electromagnets installed in a 6.3 km circular tunnel. However, during the development phase Wilson already had in mind the installation in the same tunnel of a higher energy accelerator based on superconducting magnets. The tunnel and the surface service buildings were designed and constructed to accommodate such a later addition. Operations of the Main Ring were initiated in 1972 and a superconducting magnet development program was launched a few years later, in 1976. This effort was spearheaded by Alvin Tollestrup and within a few years was producing accelerator quality superconducting magnets [5]. Also in 1976 C. Rubbia, D. Cline, and P. McIntyre proposed the construction of a proton-antiproton collider at either CERN or Fermilab [6]. They argued that a luminosity of $1\times10^{29}$ cm$^{-2}$sec$^{-1}$ would be sufficient for discovery of the W and Z bosons, and that this luminosity could be supported by a source capable of producing $1\times10^{11}$ antiprotons per day.

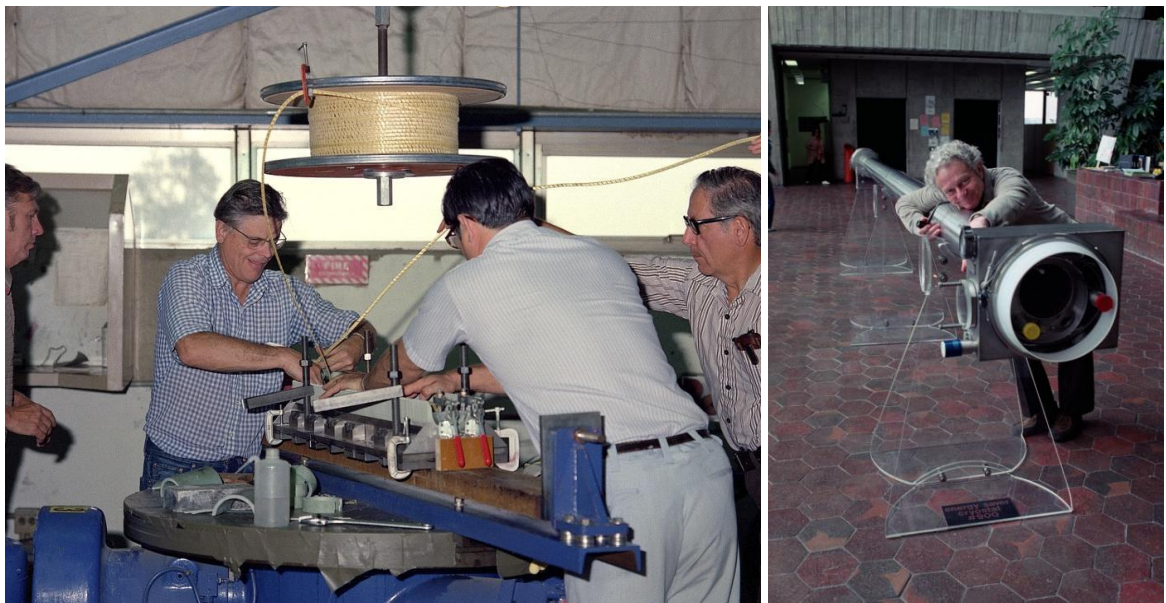

Figure 1: The first two directors of Fermilab. Bob Wilson (right) enthusiastically engages in the winding of a superconducting coil. Leon Lederman (left) expresses his affection for a Tevatron cryostat.

In 1978 CERN initiated construction of the first proton-antiproton collider based on the SPS machine and utilizing the recently developed stochastic cooling technique of Simon van der Meer. The CERN collider commenced operations in 1981 at an energy of 630 GeV (center-of-mass), leading to the discovery of the W and Z bosons in 1983. Also in 1978 the new Fermilab Director, Leon Lederman, called together all proponents of various colliding beam schemes then under development for what became known as the "Armistice Day Shootout" – November 11, 1978. Following this meeting all efforts were consolidated into construction of the Energy Saver/Doubler (as the Tevatron was then known) as the laboratory's highest priority, and it was decreed that proton-antiproton collisions would take place in the Tevatron. A new laboratory section, the Doubler Section, led by Rich Orr and Helen Edwards, was created to coordinate construction and commissioning of the Tevatron. The Accelerator Division was given responsibility for the design and construction of an Antiproton Source. It was recognized that while the Tevatron could not beat the CERN machine to the W and Z, ultimately the higher energy would win out.

In June 1981 Maury Tigner, of Cornell University, was asked by Lederman to chair a committee to review the plan that had been developed for the Antiproton Source and the associated estimate of achievable luminosity. The plan at that time was based on a hybrid scheme incorporating both stochastic and electron cooling. The committee reaffirmed the performance goal for the collider, a luminosity of $1\times10^{30}$cm$^{-2}$sec$^{-1}$ at a center-of-mass energy of 1800 GeV, and issued the following comment and recommendation:

- The design appears to be adequate to meet the goals for antiproton production and accumulation listed in the design report. However those goals are too modest.
- We recommend that the Laboratory re-examine the goals and develop a feasibility design commensurate with the full potential of the Main Ring-Booster combination to produce antiprotons.

This committee recommendation played a critical role in the development of the collider over the next 30 years. They essentially realized "it's the pbars, stupid" – a theme that repeats through the balance of the Tevatron lifetime. The Fermilab plan was producing fewer antiprotons per hour than the CERN complex despite starting with a much higher performance proton source. So it was back to the drawing board.

A new, all stochastic cooling, concept was developed and constructed under the leadership of John Peoples. The design featured the introduction of a "Debuncher" ring in which the large momentum spread of the antiprotons collected off the production target was reduced through a longitudinal phase space rotation prior to injection into the "Accumulator" ring. The Debuncher innovation significantly improved performance of the system and was later introduced into the antiproton facility at CERN. The new design capitalized on the full power of the Fermilab Linac, Booster and Main Ring and provided flexibility for future performance upgrades.

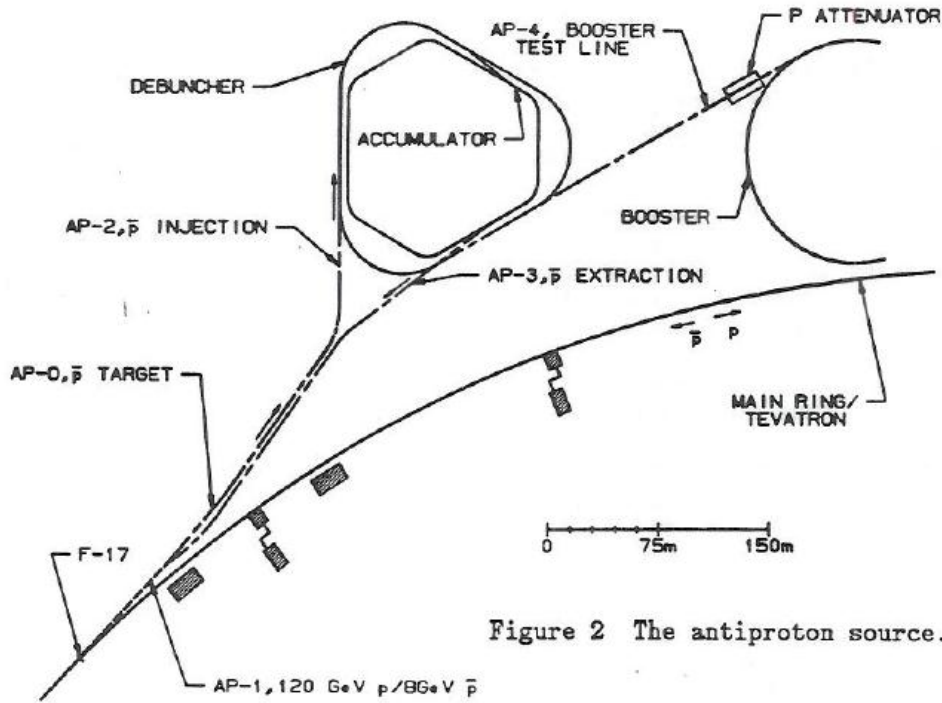

Figure 2: The Fermilab Antiproton Source design layout.

Tevatron construction was formally authorized by the Department of Energy in 1979. This was followed by authorization of the Antiproton Source and CDF detector construction in 1982. First accelerated beams were achieved in the Tevatron in July 1983 at an energy of 512 GeV. This was a tremendous achievement and largely due to the four people shown in Figure 3 (shown at the White House receiving the National Medal of Technology from President George H.W. Bush (not pictured) in 1989). Pictured from left to right are: Alvin Tollestrup who provided the leadership in developing superconducting magnets that could be used in a particle accelerator; Rich Orr who served as Head of the Double Section and so ably managed the other three (no mean feat); Dick Lundy who developed and ran the "magnet factory" at which approximately 1000 superconducting magnets were constructed on the Fermilab site; and Helen Edwards who provided the technical leadership for the design, construction, and commissioning of the Tevatron.

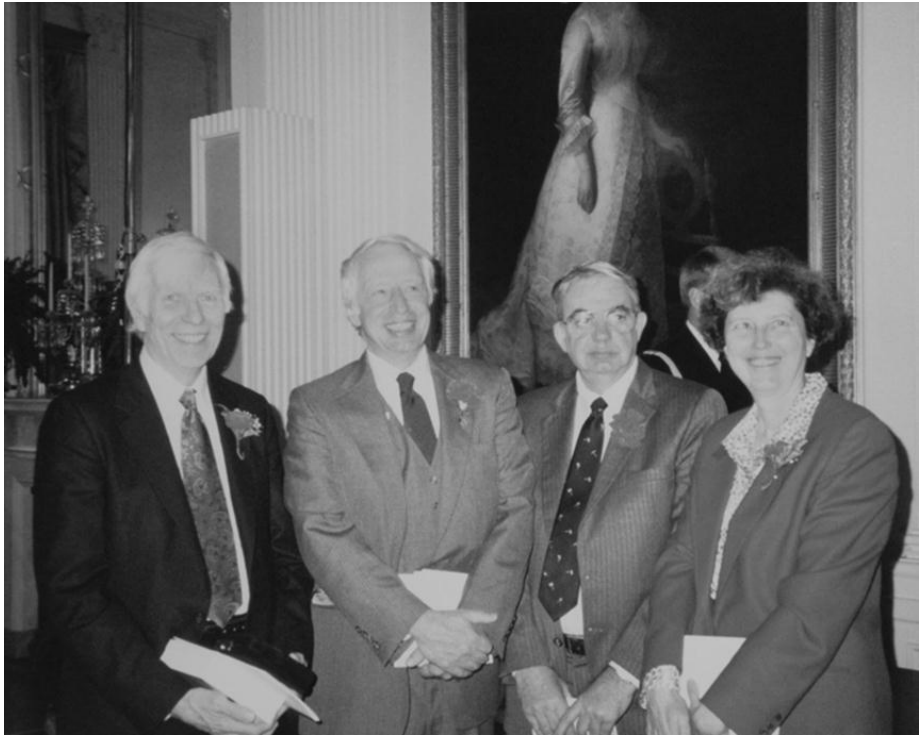

Figure 3: Alvin Tollestrup, Rich Orr, Dick Lundy, and Helen Edwards receiving the National Medal of Technology in 1989.

By 1985 the Antiproton Source had been completed and the first western hemisphere proton-antiproton collisions were observed in the partially completed CDF detector early in the morning on October 13, 1985. The collider era was now underway at an estimated luminosity of $2\times10^{25}$cm$^{-2}$sec$^{-1}$ – only five orders of magnitude to go!

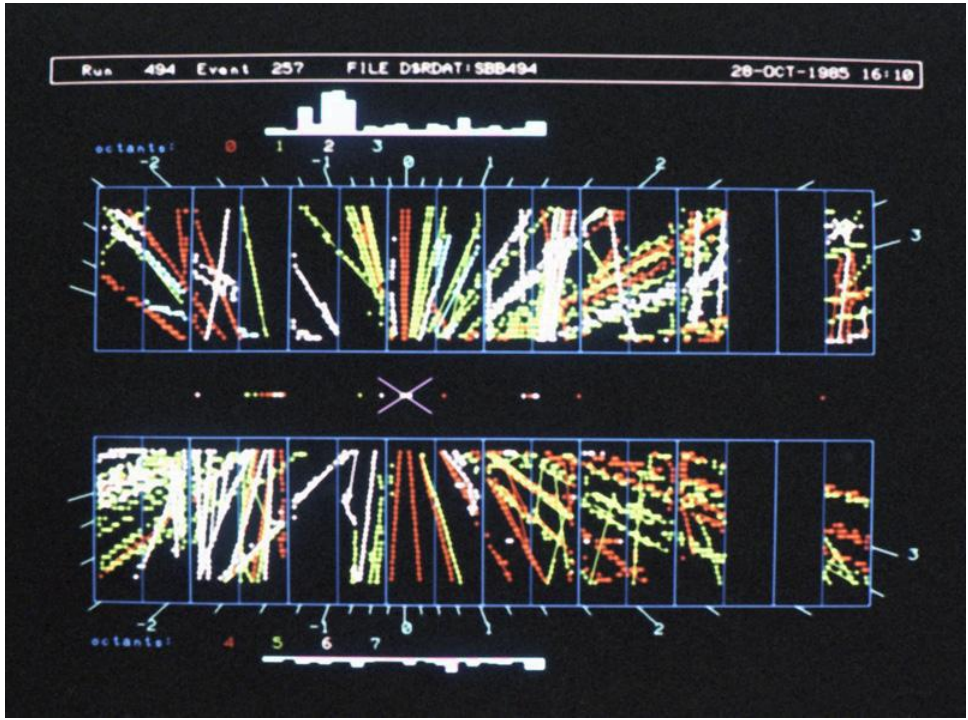

Figure 4: One of the first proton-antiproton collisions recorded in the CDF vertex detector at Fermilab on October 13, 1985

## 3. Meeting the Goals (1985-89)

Following the first observation of collisions in the fall of 1985 the Tevatron was operated for two years at an energy of 800 GeV for fixed target experiments. During this period the CDF detector was completed and modifications were made to the accelerator complex to rectify issues identified in 1985. The first extended run of the Tevatron Collider took place over several months in 1987. During this period a luminosity of $2\times10^{29}$ cm$^{-2}$sec$^{-1}$ at was achieved.

Following a nearly year-long return to fixed target operations the first extended run of the Collider took place over 1988-89 ("1988-89 Run"). In this run the design luminosity of the Tevatron was achieved for the first time. Table 1 summarizes the performance of the 1988-89 Run in comparison with the Tevatron design. The listed "actual" luminosity represents the average initial luminosity achieved over the last several months of the run. A total of 5 pb$^{-1}$ (0.005 fb$^{-1}$) was delivered to the CDF detector during this run.

Table 1: Tevatron Collider performance summary for the 1988-89 Run.

|  | **Tevatron Design** | **1988-89 Actual** |  |
|---|---|---|---|
| Energy | 1800 | 1800 | GeV |
| Protons/bunch | $6\times10^{10}$ | $7\times10^{10}$ |  |
| Antiprotons/bunch | $6\times10^{10}$ | $3\times10^{10}$ |  |
| Bunches | 3 | 6 |  |
| Luminosity | $1\times10^{30}$ | $1.6\times10^{30}$ | cm$^{-2}$sec$^{-1}$ |

## 4. Establishing the Strategy for the Future (1987-91)

Even as preparations were underway for the 1988-89 Run, planning for the next steps in performance improvement were underway. The strategy that was developed at this time, and followed for the next 15 years, can be understood in terms of the luminosity formula below.

$$L = \frac{fN_p(BN_{\bar{p}})}{2\pi(\sigma_p^2 + \sigma_{\bar{p}}^2)} H\left(\frac{\sigma_z}{\beta^*}\right) = \frac{3\gamma fN_p(BN_{\bar{p}})}{\beta^*(\varepsilon_p + \varepsilon_{\bar{p}})} H\left(\frac{\sigma_z}{\beta^*}\right)$$

The luminosity formula is written in this form to emphasize the limitations in a proton-antiproton collider. $f$ is the revolution frequency of the Tevatron, $N_p$ is the number of protons per bunch, $N_{\bar{p}}$ is the number of antiprotons in a bunch, $B$ is the number of bunches (per beam), $\sigma_p$ and $\sigma_{\bar{p}}$ are the rms transverse beam sizes of the protons and antiprotons at the interaction point, $\gamma$ is the relativistic factor, $\beta^*$ is the lattice beta function at the interaction point (assumed equal in the two transverse directions), $\varepsilon_p$ and $\varepsilon_{\bar{p}}$ are the 95% normalized emmitances of the protons and antiprotons, and $H$ is a form factor dependent on the ratio of the bunch length to the beta function ($H \approx 1$ for $\sigma_z \ll \beta^*$). The luminosity can be seen to be dependent on three primary factors: the total number of antiprotons in the Tevatron, $BN_{\bar{p}}$; the phase space density of the proton bunches, $N_p/\varepsilon$; and the beta function at the interaction point, $\beta^*$. The second of these, the proton phase space density, is in turn directly proportional to the beam-beam tune shift per crossing. The total beam-beam tune shift is this number times the number of crossings (*2B* in the case of non-separated beams). The basic strategy followed for improvement of Tevatron Collider performance over the subsequent two decades was based on three elements:
1) Provide more antiprotons to the Tevatron
2) Eliminate the beam-beam tune shift as a consideration, and then increase the proton bunch density
3) Decrease $\beta^*$ at the interaction points

In practice the improvements implemented along these lines came in two phases, designated Run I and Run II.

### 4.1. Strategy for Run I

At the time of the 1988-89 run the Run I strategy was already in place. This strategy consisted of the following major elements:

1) Helical orbits
   A total of twenty-two electrostatic separators were introduced into the Tevatron to create helically separated orbits for the protons and antiprotons. Each separator was 3 meters in length and capable of producing an electric field of ±300 kV /5 cm. When energized, head-on collisions would only occur in two locations – the interaction regions housing the CDF and D0 detectors. As a result of the implementation of helical orbits the beam-beam tune shift ceased to be a limiting issue for the balance of the Tevatron's lifetime, despite subsequent increases in proton beam intensity.

2) New low-β systems (B0 and D0)
   The 1988-89 run was conducted with only the CDF detector operational. The interaction region optics at CDF was not optically matched. Prior to Run I two identical low-β systems, providing matched optics at both CDF and D0, were installed. The β function at CDF was reduced by about a factor of two.

3) 400 MeV linac upgrade
   The energy of the linac was doubled, from 200 MeV to 400 MeV. This improved performance in the 8 GeV Booster through the reduction of space-charge forces at injection. The result was an increase in the number of protons delivered from the Booster and a higher proton beam density. This allowed the delivery of more protons to the antiproton production target, and higher intensity proton beams to the Tevatron.

4) Antiproton Source improvements
   Improvements were made to the stochastic cooling systems in the Antiproton Source: transverse cooling was introduced into the Debuncher while the system bandwidth in the Accumulator was increased. Both improvements supported a higher antiproton accumulation rate in the presence of increase proton flux onto the antiproton production target.

5) Cryogenic Cold Compressors

Cryogenic cold compressors were introduced into the Tevatron in order to lower the operating temperature by 0.5 K. The goal was to raise the Tevatron operating energy from 900 to 1000 GeV (thereby finally justifying its name!).

This set of improvements was already in the advanced planning or fabrication phase at the time of the 1988-89 run, and formed the basis for Run I. The performance goal for Run I was established as a luminosity of $1\times10^{31}$ cm$^{-2}$sec$^{-1}$, a factor of ten beyond the original Tevatron goal.

## 4.2. Strategy for Run II

And we were already thinking beyond. In the spring of 1987 Helen Edwards asked Gerry Dugan, John Marriner, and myself to look at what new construction might be required to support a luminosity of $5\times10^{31}$ cm$^{-2}$sec$^{-1}$. We developed a variety of options and wrote a report (Fermilab TM-1491; see Figure 5 for the introductory page). One can already see here the problems we were grappling with in increasing the luminosity – the need to be able to make antiprotons faster; and the requirement that we be able to store the requisite number of antiprotons somewhere.

REPORT OF THE NEW RINGS STUDY GROUP

S.D. Holmes, G. Dugan, J. Marriner
October 19, 1987

I. INTRODUCTION AND SCOPE

We have looked into the need for and possibility of constructing new accelerators at Fermilab in support of the proposed upgrade of the proton-antiproton collider to a luminosity of $5\times10^{31}$ cm$^{-2}$s$^{-1}$. The upgrade is based on running multiple batches (144 in the latest scheme) on separated orbits in the Tevatron. Beams containing on the order of $6\times10^{12}$ protons and $3\times10^{12}$ antiprotons with transverse emittances of $12\pi$ mm-mr are required. It is unclear what the required regeneration rate for collider beams will be, but it will probably lie in the range 12-24 hours. It is expected that luminosity degradation due to emittance dilution will be much more significant than luminosity degradation due to beam loss.

Antiproton economics represent one of the outstanding problems to be solved in the proposed upgrade. There are two shortcomings in the present antiproton source: 1) the accumulation rate is not high enough to support the upgrade; and 2) nowhere within the complex is there a place in which $3\times10^{12}$ antiprotons can be stored. With two interaction regions each running with an

Figure 5: First page from Fermilab TM-1491, which outlined options for new accelerators at Fermilab to support a luminosity of $5\times10^{31}$ cm$^{-2}$sec$^{-1}$.

TM-1491 defined a number of options, one of which was selected as most promising over the next two years – a "New Main Ring" which was ultimately named the Main Injector. More interesting was the suggestion of a new ring for "antiproton storage and recovery", which ultimately became the Recycler. The Main Injector was designed as a faster cycling, larger aperture, version of the Main Ring. The goal was to significantly improve the antiproton production rate and to allow more efficient transmission of stored antiprotons from the Antiproton Accumulator into the Tevatron. A valuable by-product was removing backgrounds in the D0 detector due to the traversal of their calorimeter by the Main Ring! The Main Injector was deemed sufficient by itself for the generation of a luminosity of $5\times10^{31}$ cm$^{-2}$sec$^{-1}$. The Recycler was not (yet) part of the plan as we entered the 1990s.

## 5. Achieving the Full Potential of the Tevatron (1992-2011)

Run I was initiated in August 1992 and completed in two phases ending in February 1996. In parallel the Main Injector was approved for construction, with physical construction starting in the spring of 1993. Figure 6 shows the groundbreaking ceremony on a snowy day in March of 1993. From the left are John People (now the Fermilab Director), Representative Dennis Hastert, Senator Carol Moseley Braun, Senator Paul Simon, and DOE Director for High Energy and Nuclear Physics Wilmot Hess.

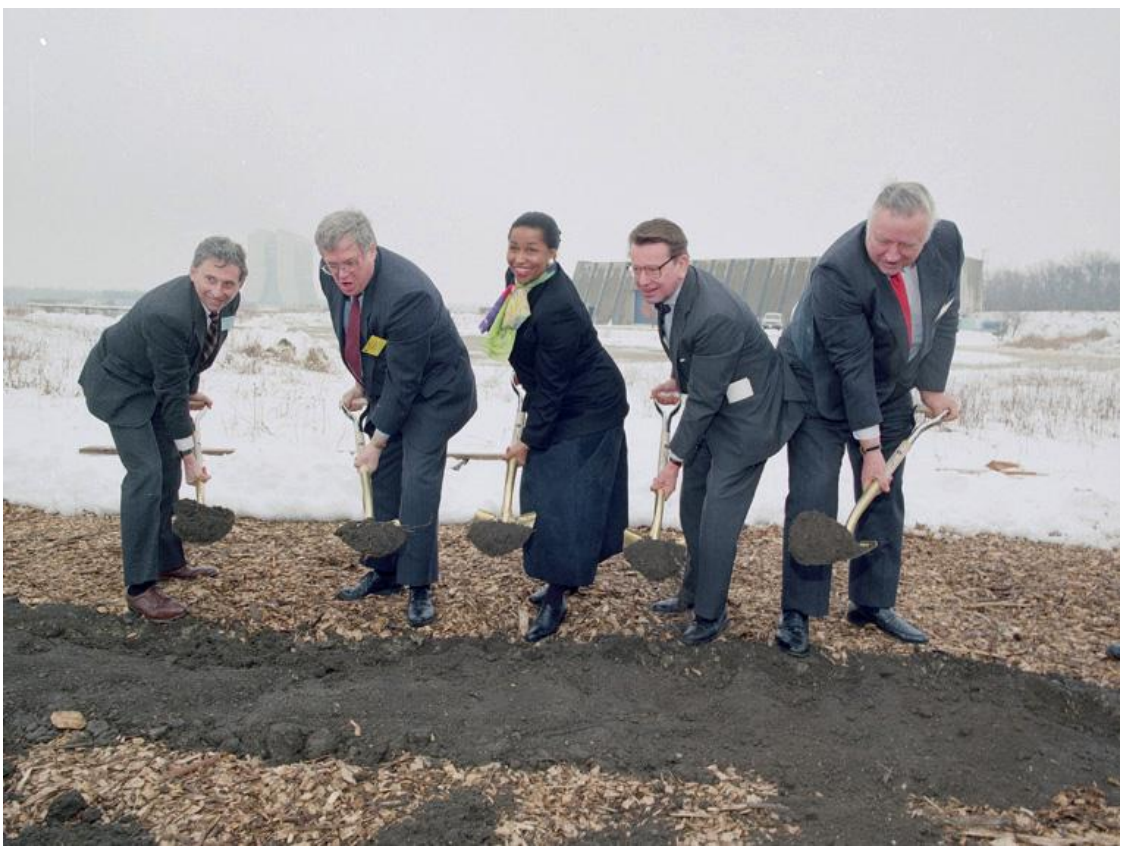

Figure 6: Main Injector ground breaking, March 1993

As construction proceeded on the Main Injector it became evident that the project was under running its budget. As much as I would like to attribute this to excellent management a more likely explanation was the prevailing economic conditions at the time, which were very conducive to low bids on major construction and fabrication contracts. This allowed the idea of a new antiproton storage ring to resurface, but with a brilliant twist. Two young scientists, Bill Foster and Gerry Jackson, came to me and John Peoples with an idea – let's build an 8 GeV antiproton storage ring (called the Recycler) in the Main Injector tunnel utilizing the Main Injector contingency funds. Their idea to make it affordable was to build it out of permanent magnets. No such application of permanent magnets on this scale had ever been undertaken anywhere, but Bill and Gerry argued forcefully. We did two things – we decided to construct the transfer line from the 8 GeV Booster to the Main Injector out of permanent magnets to gain experience and confidence, and in parallel started writing a Conceptual Design Report for the Recycler. The culmination of these activities was an Engineering Change Request (ECR) within the Main Injector Project for the utilization of contingency funds to support Recycler construction. The ERC was approved by DOE in February of 1997 (Figure 7). The luminosity goal for the Main Injector Project was raised to $8\times10^{31}$ cm$^{-2}$sec$^{-1}$ although our calculations indicated we could get as high as $2\times10^{32}$ cm$^{-2}$sec$^{-1}$ if either antiproton recycling or electron cooling were made to work.

Figure 7: The approved Engineering Change Request incorporating the Recycler into the Main Injector Project.

The Main Injector and Recycler were completed in 1999 with a short (and final) Tevatron fixed target run immediately following completion. Commissioning activities for Collider Run II were initiated in early 2000 and Run II was formally inaugurated in June 2001. A number of difficulties were experienced (and widely publicized) during the initial stages of Run II. These difficulties led to the establishment of the Run II Luminosity Upgrade Campaign in early 2002. It is important to acknowledge the enlightened attitude of the Department of Energy in supporting this campaign – in fact Dan Lehman, director of the Office of Project Assessment within the DOE Office of Science, came up with the "campaign" terminology. While the DOE certainly reviewed us a lot, they also gave us the freedom to conduct this campaign outside the normal confines of a construction project – probably the only management approach that would have allowed success.

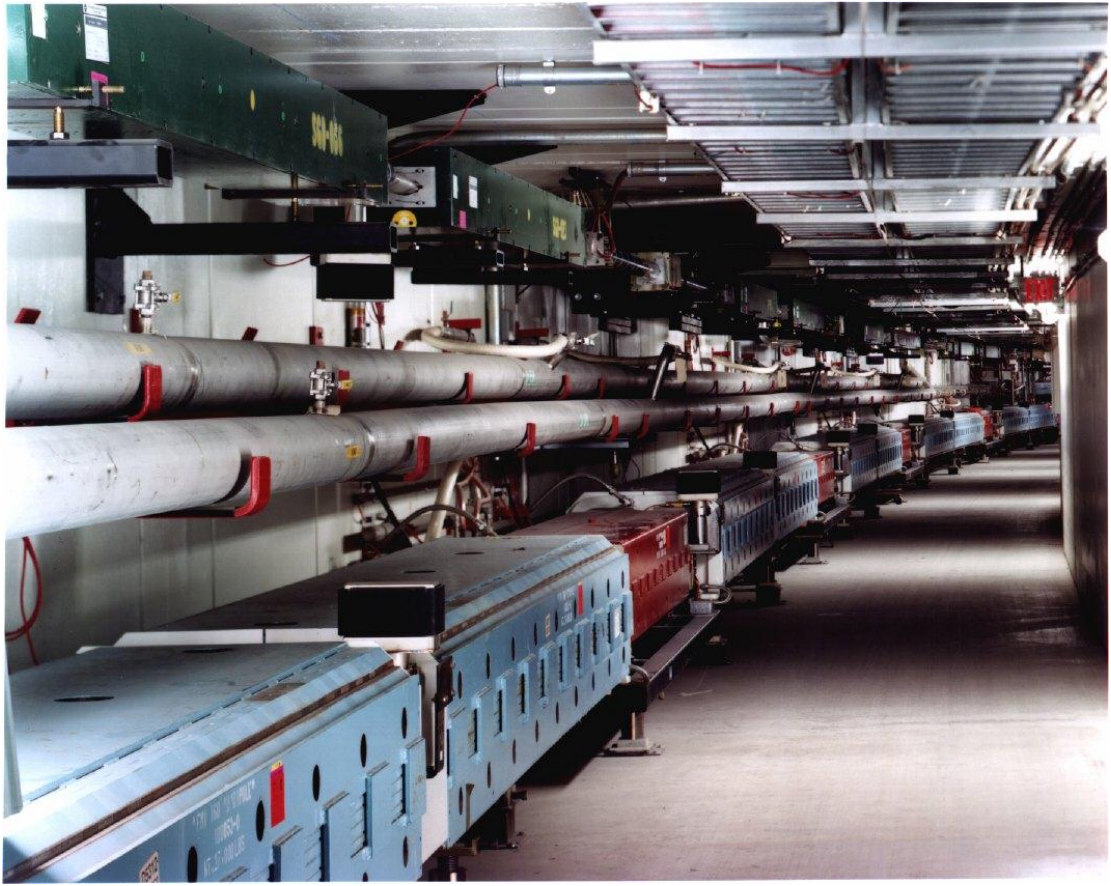

Figure 8: The completed Main Injector (bottom) and Recycler (top) within their shared enclosure. Note the absence of any electrical or water connections to the Recycler magnets!

The Run II campaign was based on literally dozens of 10-15% improvements. New ground was broken in accelerator physics in a number of areas including longitudinal "slip stacking" of proton beams in the Main Injector and the utilization of a "Tevatron Electron Lens" (a low energy, high current, electron beam that could be modulated at >2.5 MHz) to modify the tunes of individual bunches within the Tevatron. By the summer of 2004 we were routinely achieving luminosities above the initial Main Injector goal of $5\times10^{31}$ cm$^{-2}$sec$^{-1}$. We continued to introduce improvements while consistently emphasizing that there was no "silver bullet" for getting the luminosity up to design values associated with the Recycler. However, this was not quite true – there was one silver bullet in our arsenal.

Electron cooling had been under development during the late stages of Recycler construction and into the early days or Run II. It had not been included in the Recycler baseline because cooling of beams at this energy (8 GeV) had never been attempted before and there was no guarantee of success. (The Recycler baseline included a less powerful stochastic cooling system.) The electron cooling project was based on a 4.3 MeV Pelletron – an electrostatic generator capable of generating a several hundred mA, dc, beam. The project was led by Sergei Nagaitsev. Not only was Sergei a very talented scientist, but he also had a great understanding of human psychology. He spent the entire development period warning management that it was far from a sure thing that this would work at all. However, when electron cooling of antiprotons in the Recycler was first attempted in the summer of 2005 it worked within a number of hours (Figure 9).

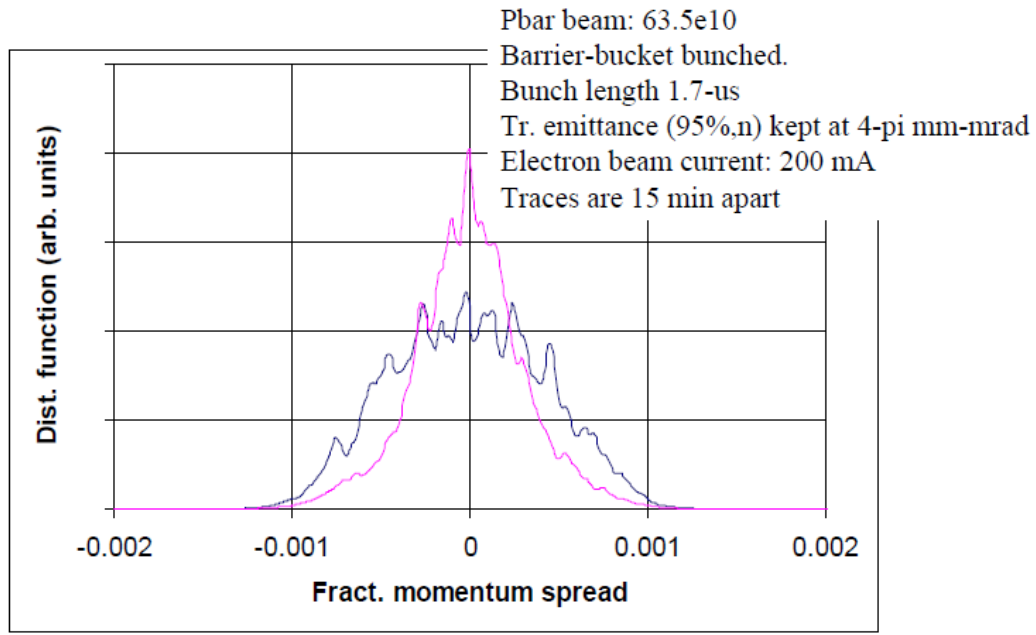

Figure 9: First demonstration of the electron cooling of 8 GeV antiprotons in the Recycler. Two momentum scans separated by 15 minutes are shown.

Electron cooling was a huge success and ultimately was worth probably a factor of ~3 in luminosity – no store exceeding $1\times10^{32}$ cm$^{-2}$sec$^{-1}$ was ever achieved without electron cooling. Run II eventually covered a decade, from June 2001 through September 2011. A total of 12 fb$^{-1}$ was delivered to both the CDF and D0 experiments. A summary of the achievements of Runs I and II are given in Table 2.

Table 2: Tevatron Collider performance summary for Run I and Run II.

|  | **Tevatron Design** | **Run I** | **Run II** |  |
|---|---|---|---|---|
| Energy | 1800 | 1800 | 1960 | GeV |
| Protons/bunch | $6\times10^{10}$ | $23\times10^{10}$ | $29\times10^{10}$ |  |
| Antiprotons/bunch | $6\times10^{10}$ | $6\times10^{10}$ | $8\times10^{10}$ |  |
| Bunches | 3 | 6 | 36 |  |
| Luminosity | $1\times10^{30}$ | $16\times10^{30}$ | $340\times10^{30}$ | cm$^{-2}$sec$^{-1}$ |
| Dates |  | 1992-96 | 2001-2011 |  |
| Integrated Luminosity |  | 0.2 | 12 | fb$^{-1}$ |

## 6. Appreciating the Tevatron

The history of the Tevatron Collider is shown graphically in Figure 10 – the initial luminosity of every store is presented here. One measure of progress is that when we started we measured performance in terms of $nb^{-1}$ accumulated; in the 1990's in terms of $pb^{-1}$; and at the end in $fb^{-1}$.

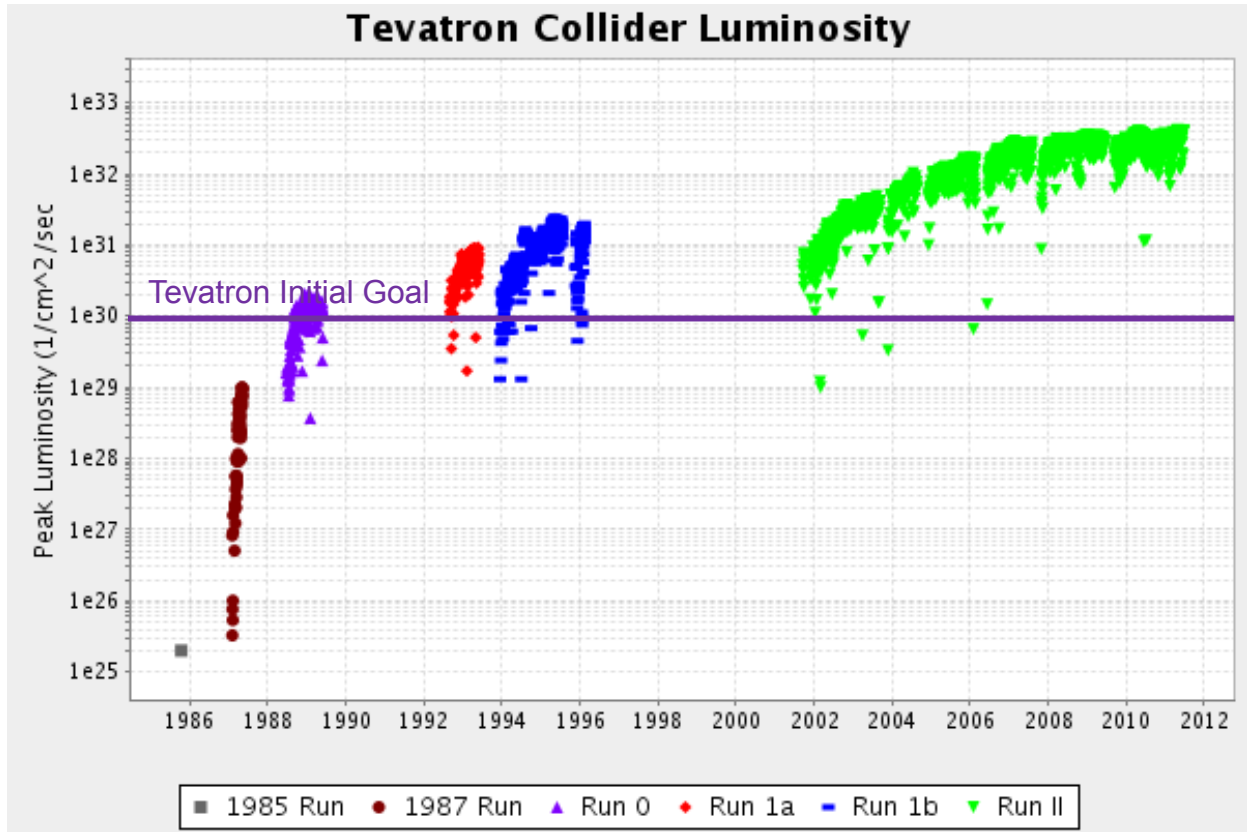

Figure 10: The luminosity history of the Tevatron Collider

The Tevatron ultimately exceeded its initial luminosity goal by a factor of 400. This represents an annualize performance growth of 35%, maintained over 20 years – far outperforming the stock market over the same period! On October 4, 2005 the Tevatron became the highest luminosity hadron collider ever operated, surpassing the CERN ISR (Intersecting Storage Rings, a proton-proton collider based on dc beams) with a luminosity in excess of $140 \times 10^{30}$ $cm^{-2}sec^{-1}$. The success of the Tevatron made possible the next generation of hadron colliders – HERA and RHIC – and ultimately set the stage for the LHC. On April 22, 2011 the newly constructed LHC surpassed the Tevatron in luminosity for the first time, achieving a luminosity of $470 \times 10^{30}$ $cm^{-2}sec^{-1}$. This is a cause for celebration as it marks continued progress in the development of the superconducting high energy collider as an enabling facility for elementary particle physics research.

## 7. Final Thoughts

I have prepared this presentation as a representative of the thousands of people who achieved great success by working together with great enthusiasm and dedication. Scientists, engineers, designers, drafters, programmers, technicians, secretaries, buyers, contract administrators, safety professionals, lawyers, financial administrators, janitors, cafeteria workers – all these people put their hearts into the Tevatron and made it what it is. It was a great run and we all had fun doing it.

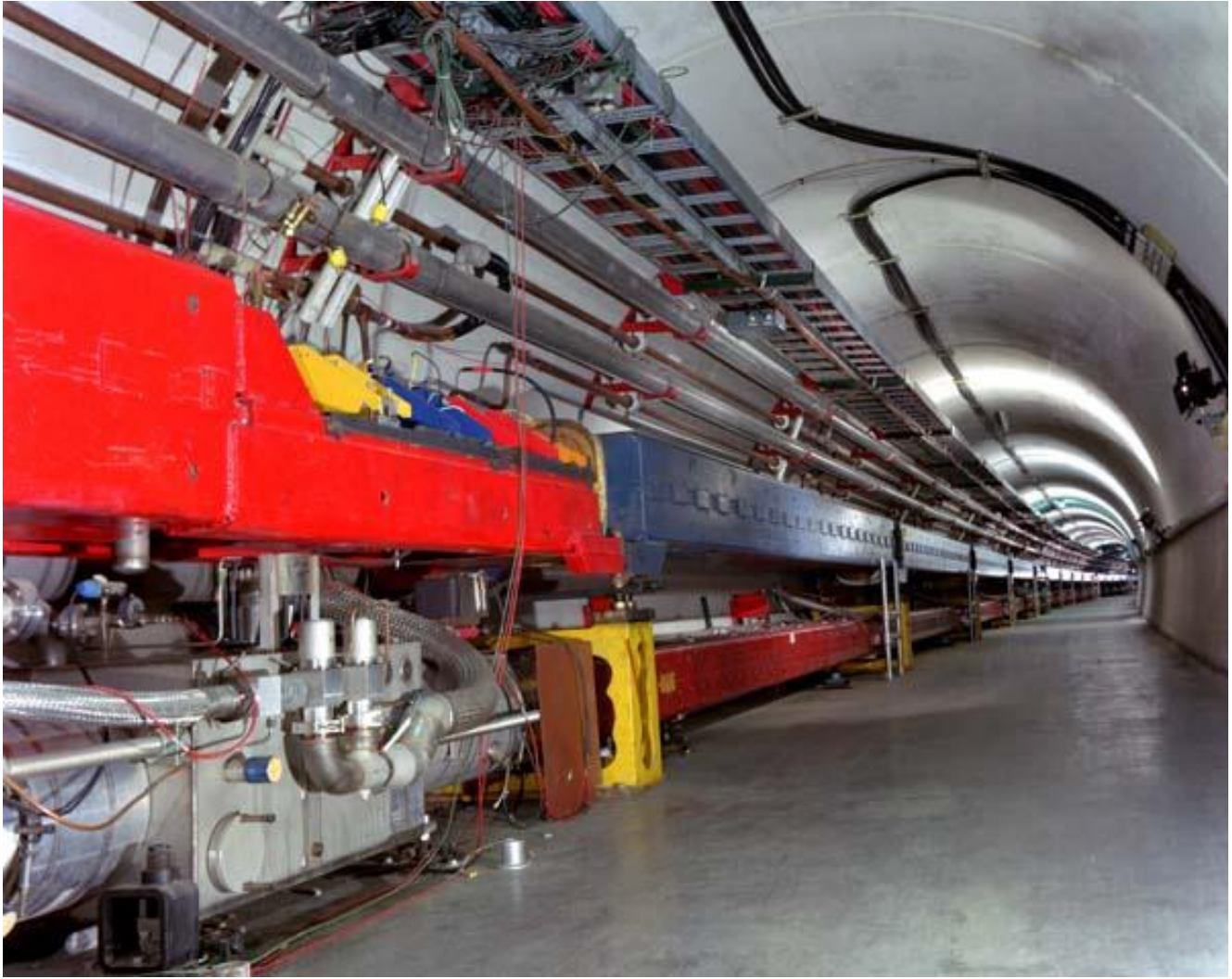

Figure 11: The Tevatron situated in its 6.28 (2π) km tunnel, directly underneath the original Main Ring.